\documentclass[aps,prl,floatfix,epsfig,twocolumn,showpacs,preprintnumbers]{revtex4}
\usepackage{amssymb}
\usepackage{amsmath}
\usepackage{graphicx}
\usepackage{epstopdf}
\usepackage{dcolumn}
\usepackage{bm}
\usepackage{mathrsfs}
\usepackage[pdfstartview=FitH, CJKbookmarks=true, bookmarksnumbered=true, bookmarksopen=true, pdfborderstyle={/S/U}, pdfborder={0 0 1}, colorlinks=true, linkcolor=blue, anchorcolor=green, citecolor=blue]{hyperref}

\DeclareMathAlphabet      {\mathbf}{OT1}{cmr}{bx}{n}
\begin{document}

\title{Calculated Spin Fluctuational Pairing Interaction in HgBa$_{2}$CuO$%
_{4}$ using LDA+FLEX Method}
\author{Griffin Heier and Sergey Y. Savrasov}
\affiliation{Department of Physics, University of California, Davis, CA 95616, USA}

\begin{abstract}
A combination of density functional theory in its local density
approximation (LDA) with k- and $\omega $ dependent self--energy found from
fluctuational--exchange--type random phase approximation (FLEX--RPA) is
utilized here to study superconducting pairing interaction in a prototype
cuprate superconductor HgBa$_{2}$CuO$_{4}$. Although, FLEX--RPA methodology
have been widely applied in the past to unconventional superconductors,
previous studies were mostly based on tight--binding derived minimal
Hamiltonians, while the approach presented here deals directly with the
first principle electronic structure calculation of the studied material
where spin and charge susceptibilities are evaluated for a correlated subset
of the electronic Hilbert space as it is done in popular LDA+U and LDA+DMFT
methods. Based on our numerically extracted pairing interaction among the
Fermi surface electrons we exactly diagonalize a linearized BCS gap
equation, whose highest eigenstate is expectantly found corresponding to $%
d_{x^{2}-y^{2}}$ symmetry for a wide range of on-site Coulomb repulsions U
and dopings that we treat using virtual crystal approximation. Calculated
normal state self--energies show a weak k- and strong frequency dependence
with particularly large electronic mass enhancement in the vicinity of spin
density wave instability. Although the results presented here do not bring
any surprisingly new physics to this very old problem, our approach is an
attempt to establish the numerical procedure to evaluate material specific
coupling constant $\lambda $ for high T$_{c}$ superconductors without
reliance on tight--binding approximations of their electronic structures.
\end{abstract}

\date{\today }
\maketitle

\section{I. Introduction.}

Shortly before the discovery of high--temperature superconductivity in
cuprates in 1986\cite{1986}, two seminal works \cite{Scalapino,Varma} have
been published in an attempt to understand properties of heavy fermion
superconductors by the pairing of their Fermi surface electrons mediated by
strong (anti)ferromagnetic spin fluctuations which can lead to symmetries of
the superconducting state of angular momenta higher than zero. Although such
random--phase approximation (RPA) based calculations deemed oversimplified,
the divergency of spin susceptibility in the vicinity of the magnetic, spin
density wave (SDW) type instability due to the Fermi surface nesting is a
common feature of many unconventional superconductors which this method
naturally incorporates. The approach took off right after doped La$_{2}$CuO$%
_{4}$ was shown to superconduct at 33K\cite{BM} and has been applied since
then to study unconventional superconductivity phenomenon \cite{Review} in a
great variety of materials, such as cuprates\cite{Shimahara,Pines,Arita,Ueda}%
, ruthenates\cite{Mazin,Takimoto}, cobaltates\cite{Co}, ironates \cite%
{Ikeda,JXLi,Schm,Hirsch}, heavy fermion\cite{Takimoto2,Tada} systems, and
most recently, nickelates\cite{Kuroki,Dagotto}.

To date, most of these applications however utilize simple few--orbital
models where the hopping integrals are extracted from density functional
based calculations using such popular approximations as Local Density
Approximation (LDA)\cite{DFT}, and these parameters are subsequently treated
as the input to the Hubbard--type model Hamiltonians. The latter is then
solved by an available many--body technique, such, for example, as the
Fluctuational Exchange Approximation (FLEX) \cite{FLEX}. FLEX is a
diagrammatic approach that includes particle--hole ladders and bubbles as
well as particle--particle ladder diagrams while the RPA neglects the latter
contribution. However, it was found to be sufficiently small \cite{Muller}
at least for the problem of paramagnons\cite{Paramagnons,BerkSchrieffer}
where the most divergent terms are given by the particle--hole ladders.

Many past studies of strongly correlated systems have been performed using
the RPA and FLEX \cite{FLEX-Review} including the proposals to combine it
with density functional electronic structure calculations \cite{LDA++}. More
recently developed combination of LDA with Dynamical Mean Field Theory
(LDA+DMFT)\cite{LDA+DMFT} sometimes utilizes the local FLEX approximation to
solve corresponding impurity problem during the self--consistent solution of
the DMFT equations. A further combination of FLEX and DMFT was also proposed
recently and has resulted in reproducing a doping dependence of critical
temperature seen in cuprates \cite{FLEX+DMFT}. More rigorous Quantum Monte
Carlo based simulations provide further extensions to this approach\cite%
{Maier,Maier2}.

We have recently described an implementation of the LDA+FLEX(RPA)\cite%
{LDA+FLEX} approach\ using the method of projectors which allows to evaluate
dynamical susceptibilities of the electrons in a Hilbert space restricted by
correlated orbitals only. This is very similar to how it is done in such
popular electronic structure techniques as LDA+U\cite{LDA+U,LDA+U-Review}
and LDA+DMFT\cite{LDA+DMFT}. The projector formalism tremendously simplifies
the numerics and allows to incorporate $\mathbf{k}$-- and $\omega $
dependent self--energies of correlated electrons straight into the LDA\
electronic structure calculation. Our applications to V and Pd \cite%
{LDA+FLEX} have, in particular, showed that the d--electron self--energies
in these materials are remarkably k--independent which justifies the use of
local self--energy approximations, such as DMFT.

Here, we extend the projector based LDA+FLEX approach to evaluate
superconducting pairing interactions describing the scattering of the Cooper
pairs at the Fermi surface in a realistic material framework. We utilize
density functional calculation of the electronic energy bands and wave
functions for HgBa$_{2}$CuO$_{4}$, a prototype single--layer cuprate whose
superconducting $T_{c}$ was reported to be 94K\cite{Hg-Supra}. Based on our
numerically evaluated pairing function we exactly diagonalize a linearized
BCS gap equation on a three dimensional k--grid of points in the\ Brillouin
Zone. The extracted highest (in value) eigenstate from this procedure is
unsurprisingly found to correspond to $d_{x^{2}-y^{2}}$ symmetry for a wide
range of on--site Coulomb repulsions $U$ and dopings that we scan during our
simulations. The corresponding maximum eigenvalue $\lambda _{\max \text{ }}$%
represents a coupling constant similar to the parameter $\lambda _{e-p}$ of
the electron--phonon (e-p) theory of superconductivity. Our primary goal
here is to establish a numerical procedure for the material specific
evaluation of this coupling constant that can hopefully be helpful in future
findings of the materials with high $T_{c}.$We however found $\lambda _{\max 
\text{ }}$ to be very sensitive to the values of $U$ used in our calculation
once we approach the region of antiferromagnetic instability. The same is
seen in our calculated normal state self--energies which were found to show
a weak $\mathbf{k-}$ and strong frequency dependence with particularly large
electronic mass renormalization $m^{\ast }/m_{LDA}=1+\lambda _{sf}$ in the
proximity to SDW. The evaluated renormalized coupling constant $\lambda
_{eff}=\lambda _{\max \text{ }}/(1+\lambda _{sf})$ is found to be modest and
incapable to deliver the high $T_{c}$ values unless we tune $U$ to be close
to SDW. Using the available experimental constraints on the values of $%
\lambda _{sf}\lesssim 3,$ we find $\lambda _{eff}\lesssim 0.4$ and the BCS\ $%
T_{c}\lesssim 30K.$ Despite it looks like an underestimation, we think the
approach opens up better opportunities to find material specific dependence
of the $T_{c}$ in unconventional superconductors without reliance on
tight--binding approximations of their electronic structures.

Our paper is organized as follows: In Section II we summarize the approach
to evaluate the pairing interaction using the LDA+FLEX formalism. In Section
III we discuss our results of exact diagonalization of the linearized BCS
equation and correspondingly extracted superconducting energy gaps and the
eigenvalues as a function of $U$ and doping. We also present our results for
correlated electronic structure in HgBa$_{2}$CuO$_{4}$ in the normal state,
the calculated mass enhancement, the effective coupling constant $\lambda
_{eff}$ and finally give some estimates for the $T_{c}$. Section V is the
conclusion.

\section{II. Method}

\subsection{a. Superconducting Pairing Interaction from LDA+FLEX.}

Our assumption here is that a general spin--dependent interaction is
operating between the electrons at the Fermi surface%
\begin{equation}
K^{\nu _{1}\nu _{2}\nu _{3}\nu _{4}}(\mathbf{r}_{1},\mathbf{r}_{2},\mathbf{r}%
_{3},\mathbf{r}_{4}).  \label{EQ-K}
\end{equation}%
Here for the sake of numerical simplicity we make one important
approximation to consider this interaction as static and operating between
the electrons only in the close proximity to the Fermi energy exactly as the
BCS\ theory assumes. The inclusion of its frequency dependence is of course
possible and has been done previously in many model calculations but we
postpone such implementation for real materials for the future.

For the non--relativistic formulation that is adopted here, due to full
rotational invariance of the spin space, the actual dependence of this
interaction on spin indexes appears to be the following 
\begin{equation*}
K^{\nu _{1}\nu _{2}\nu _{3}\nu _{4}}=\frac{1}{2}\delta _{\nu _{1}\nu
_{3}}\delta _{\nu _{2}\nu _{4}}K^{c}-\frac{1}{2}\mathbf{\sigma }_{\nu
_{1}\nu _{3}}\mathbf{\sigma }_{\nu _{2}\nu _{4}}K^{s},
\end{equation*}%
where the interactions $K^{c}$ and $K^{s}$ are due to charge and spin
degrees of freedom, and $\mathbf{\sigma }$ are the Pauli matrices.
Transformation to singlet--triplet representation is performed using the
eigenvectors $A_{\nu _{1}\nu _{2}}^{SS_{z}}$of the product for two spin
operators which leads us to consider the interactions for the singlet ($%
S=0,S_{z}=0)$ and triplet\ $(S=1,S_{z}=-1,0,+1)$ states separately%
\begin{eqnarray*}
K^{(S^{\prime }S_{z}^{\prime }SS_{z})} &=&\sum_{\nu _{1}\nu _{2}\nu _{3}\nu
_{4}}A_{\nu _{1}\nu _{2}}^{S^{\prime }S_{z}^{\prime }}K^{\nu _{1}\nu _{2}\nu
_{3}\nu _{4}}A_{\nu _{3}\nu _{4}}^{SS_{z}} \\
&=&\delta _{S^{\prime }S}\delta _{S_{z}^{\prime }S_{z}}K^{(S)},
\end{eqnarray*}%
where $K^{(S)}=\frac{1}{2}K^{c}-\frac{1}{2}E_{S}K^{s},$ and $%
E_{S=0}=-3,E_{S=1}=+1$ are the eigenvalues for the spin product operators.

We next introduce the matrix elements of scattering between the Cooper pair
wave functions $\Psi _{\mathbf{k}j,SS_{z}}^{(\nu _{1}\nu _{2})}(\mathbf{r}%
_{1},\mathbf{r}_{2})$ which are proper antisymmetric combinations of the
electronic wave functions with their Fermi momenta $\mathbf{k}$ and $-%
\mathbf{k}$ in a given energy band labeled by index $j.$ In the
singlet--triplet representation these matrix elements are diagonal with
respect to the spin indexes and do not depend on $S_{z}$

\begin{equation}
\sum_{\nu _{1}\nu _{2}\nu _{3}\nu _{4}}\langle \Psi _{\mathbf{k}%
j,SS_{z}}^{(\nu _{1}\nu _{2})}|K^{\nu _{1}\nu _{2}\nu _{3}\nu _{4}}|\Psi _{%
\mathbf{k}^{\prime }j^{\prime },S^{\prime }S_{z}^{\prime }}^{(\nu _{3}\nu
_{4})}\rangle =\delta _{S^{\prime }S}\delta _{S_{z}^{\prime }S_{z}}M_{%
\mathbf{k}j\mathbf{k}^{\prime }j^{\prime }}^{(S)}.  \label{EQ-M}
\end{equation}%
Since one--electron wave functions forming the Cooper pairs should obey the
Bloch theorem, the integration in the matrix elements can be reduced to the
integration over a single unit cell which leads us to consider the paring
interaction in terms of its lattice Fourier transforms with various
combinations of $\pm \mathbf{k}$ and $\pm \mathbf{k}^{\prime }$ of the type:

\begin{eqnarray*}
&&K_{\mathbf{k,k}^{\prime }}^{(S)}(\mathbf{r}_{1},\mathbf{r}_{2},\mathbf{r}%
_{3},\mathbf{r}_{4})=\sum_{R_{1}R_{2}R_{3}R_{4}}e^{-i\mathbf{k}(\mathbf{R}%
_{1}-\mathbf{R}_{2})}e^{i\mathbf{k}^{\prime }(\mathbf{R}_{3}-\mathbf{R}%
_{4})}\times \\
&&K^{(S)}(\mathbf{r}_{1}\mathbf{-R}_{1},\mathbf{r}_{2}\mathbf{-R}_{2},%
\mathbf{r}_{3}-\mathbf{R}_{3},\mathbf{r}_{4}-\mathbf{R}_{4}).
\end{eqnarray*}%
(Due to translational periodicity one lattice sum should be omitted.)

The Cooper pair wave functions can be constructed from corresponding
single--electron states that are easily accessible in any density functional
based electronic structure calculation. However, the formidable theoretical
problem is to evaluate the pairing interaction $K^{(S)}$. Our first
approximation to this function is to assume that it operates for correlated
subset of electrons which are introduced with help of site dependent
projector operators: $\phi _{a}(\mathbf{r})=\phi _{l}(r)i^{l}Y_{lm}(\hat{r})$
of the one--electron Schroedinger equation taken with a spherically
symmetric part of the full potential. \cite{phidot}. The Hilbert space \{$a$%
\} inside the designated correlated site restricts the full orbital set by a
subset of correlated orbitals, such as those corresponding to $l=2$ for Cu.
We therefore write%
\begin{eqnarray*}
&&K_{\mathbf{k,k}^{\prime }}^{(S)}(\mathbf{r}_{1},\mathbf{r}_{2},\mathbf{r}%
_{3},\mathbf{r}_{4}) \\
&=&\sum_{a_{1}a_{2}a_{3}a_{4}}\phi _{a_{1}}(\mathbf{r}_{1})\phi _{a_{2}}(%
\mathbf{r}_{2})K_{a_{1}a_{2}a_{3}a_{4}}^{(S)}(\mathbf{k,k}^{\prime })\phi
_{a_{3}}^{\ast }(\mathbf{r}_{3})\phi _{a_{4}}^{\ast }(\mathbf{r}_{4})
\end{eqnarray*}%
Our second approximation is to adopt the LDA+FLEX(RPA)\ procedure for
evaluating the matrix $K_{a_{1}a_{2}a_{3}a_{4}}^{(S)}(\mathbf{k,k}^{\prime })
$ (static for this particular problem, but $\omega $ dependent in general)$.$
Namely, we represent it in terms of screening the on--site Coulomb
interaction matrix $I_{a_{1}a_{2}a_{3}a_{4}}$ (we drop all indexes hereafter
as this becomes just the matrix manipulation)%
\begin{equation*}
\hat{K}=\hat{I}+\hat{I}[\hat{\chi}-\frac{1}{2}\hat{\pi}]\hat{I}.
\end{equation*}%
Here the interacting susceptibility $\hat{\chi}=\hat{\pi}[\hat{1}-\hat{I}%
\hat{\pi}]^{-1}$, $\hat{\pi}$ is the non--interacting polarizability, and
the subtraction of $\frac{1}{2}\hat{\pi}$ takes care of the single bubble
diagram that appears twice in both bubble and ladder series. Remind that the
matrix $\hat{I}$ is local in space since it describes the on--site Coulomb
repulsion $U$. Due to this notion of locality, the screened matrix $%
K_{a_{1}a_{2}a_{3}a_{4}}^{(S)}(\mathbf{k,k}^{\prime })$ becomes dependent
only on $\mathbf{k}\pm \mathbf{k}^{\prime }.$

The procedure to calculate the matrix $\hat{K}$ using density functional
based electronic structure for real materials was described in details in
our previous publication \cite{LDA+FLEX}. Here we would only like to point
out that it is still a computationally demanding problem since the matrices
need to be computed for dense set of wavevectors and their frequency
dependence is also generally required. The restriction by the correlated
subset tremendously simplifies all matrix manipulations with the ladder
diagrams\ that rely on the 4--point functions scaling with the number of
atoms in the unit cell as $N_{atom}^{4}$. This is contrary to the bubble
diagrams which rely on the two--point functions scaling as $N_{atom}^{2},$
and are in the heart of such popular method as GW\cite{GW}. However, the use
of the on--site interaction in ladder diagrams allows one to express all
quantities via charge and spin susceptibilities which are the two--point
functions and allow to regain the $N_{atom}^{2}$ scaling$.$ It is still
computationally involved because the number of matrix elements for
representing the susceptibility grows as $N_{atom}^{2}N_{orb}^{4}$ where $%
N_{orb}$ is the size of complete orbital manifold per atom needed. For HgBa$%
_{2}$CuO$_{4}$, $N_{atom}=8,$ $N_{orb}=9$ for Hg,Ba,Cu ($l_{\max }\leq 2)$
and $N_{orb}=4$ for O ($l_{\max }\leq 1)$, this requires at least 4$^{2}$x9$%
^{4}$+4$^{2}$x4$^{4}$=109,072 matrix elements to be computed for each wave
vector and frequency! Often, to improve the accuracy, the number of orbitals
per each angular harmonic needs to be doubled or tripled which blows up the
matrices by another one to two orders due to $N^{4}$ scaling. The
restriction by the correlated subset greatly facilitates the calculation,
because now the matrices have to be computed for the correlated sites and
orbitals only, and for the problem at hand, 5 orbital states representing Cu
d--electrons produce only 1$^{2}$x5$^{4}$=625 matrix elements$.$

\subsection{b. Spin Fluctuational Coupling Constant}

The matrix elements $M_{\mathbf{k}j\mathbf{k}^{\prime }j^{\prime }}^{(S)}$
which scatter the Cooper pairs enter the Eliashberg gap equation for
superconducting $T_{c}.$ Numerous solutions of this equation have been
implemented in the past for the single-- and multi--orbital Hubbard models
to address the question of unconventional superconductivity in cuprates and
other systems \cite{Shimahara,Arita,Ueda,Takimoto, Ikeda}. These
implementations involve both the FLEX and more sophisticated Dynamical
Cluster Approximation (DCA)\cite{Maier,Maier2} for the pairing interaction;
they work on the imaginary Matsubara frequency axis and do not determine the
coupling constant directly, but go straight to the $T_{c}$. The
self--consistency is important as it allows to account for many effects
known from the theory of superconductivity. In particular, the mass of the
quasiparticles is known to renormalize due to the attractive paring
interaction operating at some small energy scale, set,\textit{\ e.g.},, by a
spin fluctuational energy $\omega _{sf}$. Also, the Coulomb repulsion that
operates at much larger energy scale, such as plasmon energy $\omega _{p},$
weakens the coupling of the Cooper pairs somewhat.

Establishing numerical procedure for estimating the coupling constant for
unconventional superconductors is central for understanding material
specific trends of their critical temperatures. This was earlier the case
for electron--phonon (e--p) superconductors \cite{EPI}, where, in most
cases, the solution of the gap equation is given by the momentum independent
gap function in the singlet pairing channel, that corresponds to $\lambda
_{e-p.}.$ Luckily, phonons have a well--defined cutoff frequency which is
the phonon Debye energy $\omega _{D},$and the equation for $T_{c}$ is often
treated in the BCS approximation, i.e., when the pairing between the
electrons resides only in a small energy window $\pm \omega _{D}$ around the
Fermi energy $\epsilon _{F}$, and where the matrix elements $M_{\mathbf{k}j%
\mathbf{k}^{\prime }j^{\prime }}^{(S)}$ are assumed to be constant, and zero
outside those energies. The quasiparticle mass enhancement, $m^{\ast }/m=1+$ 
$\lambda _{e-p.},$ that leads to a kink in the single--particle spectrum at
the scale $\omega _{D}$ and the effects of the Coulomb interaction that
weaken the coupling constant $\lambda _{e-p.}$ by the parameter $\mu ^{\ast
} $ (usually very small,\symbol{126}0.1, due to different energy scales $%
\omega _{D}$ vs. $\omega _{p}$ \cite{Anderson}) are taken into account by
utilizing the renormalized coupling constant $\lambda _{eff}=(\lambda
_{e-p.}-\mu ^{\ast })/(1+\lambda _{e-p})$ that determines the $T_{c}.$ With
a few empirically adjusted coefficients this gave rise to the famous
McMillan $T_{c}$ equation\cite{McMillan}. One, in principle, does not need
this simplified point of view and can proceed with the Eliashberg equation,
but in a semiquantitative way, the McMillan theory is known to work very
well.

In the following we adopt the BCS\ approximation by assuming that the
pairing occurs in a small region around the Fermi surface restricted by some
spin fluctuational frequency $\omega _{sf}.$ Although for a generally
screened electron--electron interaction there is no formal justification to
separate such small energy scale, it is known that spin fluctuations have a
characteristic energy $\omega _{sf}$ similar to phonons, and that
experimentally, in cuprates they have been seen in the range of energies
30--50 meV as peaks in imaginary spin susceptibility accessible via the
numerous neutron scattering experiments\cite{INS}. There is a famous 40 meV
resonance which is visible in the superconducting state\cite{40meV}. There
are numerous angle resolved photoemission experiments (ARPES) that show
kinks in the one--electron spectra at the same energy range\cite{ARPES-Kinks}%
. These kinks are sometimes interpreted as caused by the electron--phonon
interactions\cite{Lanzara}, but, unfortunately, the calculated values of $%
\lambda _{e-p.}$ are known to be small in the cuprates\cite%
{Savrasov-OKA,Louie}. Note also that for the undoped antiferromagnetic
cuprates, the spin wave spectra reside in the energy range of 30 meV \cite%
{SpinWaves}.

Thus, we will assume that the matrix elements $M_{\mathbf{k}j\mathbf{k}%
^{\prime }j^{\prime }}^{(S)}$ which scatter the Cooper pairs enter the
linearized BCS equation

\begin{eqnarray}
\Delta _{S}(\mathbf{k}j) &=&-\frac{1}{2}\sum_{\mathbf{k}^{\prime }j^{\prime
}\in \epsilon _{F}\pm \omega _{sf}}M_{\mathbf{k}j\mathbf{k}^{\prime
}j^{\prime }}^{(S)}\Delta _{S}(\mathbf{k}^{\prime }j^{\prime })\times  \notag
\\
&&\tanh \left( \frac{\epsilon _{\mathbf{k}^{\prime }j^{\prime }}}{2T_{c}}%
\right) /2\epsilon _{\mathbf{k}^{\prime }j^{\prime }}.  \label{EQ-BCS}
\end{eqnarray}%
where the summation over $\mathbf{k}^{\prime }j^{\prime }$ goes over the
electrons residing in a small region around the Fermi surface restricted by $%
\omega _{sf}$ .The solutions $\Delta _{S}(\mathbf{k}j)$ for $S=0$ or $1$
describe momentum dependence of superconducting energy gap and are known to
be either even or odd functions of momenta. Performing the integration over
the energy window $\epsilon _{F}\pm \omega _{sf}$ we rewrite the equation in
a form 
\begin{equation*}
-\ln \left( \frac{1.134\omega _{sf}}{T_{c}}\right) \sum_{i^{\prime }}M^{(S)}(%
\hat{k}_{i},\hat{k}_{i^{\prime }})\frac{\delta A_{i^{\prime }}}{%
|v_{i^{\prime }}|}\Delta _{S}(\hat{k}_{i^{\prime }})=\Delta _{S}(\hat{k}%
_{i}),
\end{equation*}%
Here we introduced some discretization of the Fermi surface onto small areas 
$\delta A_{i}$ with absolute values of the electronic velocities $\left\vert
v_{i}\right\vert $ whose locations are pointed by the Fermi momenta $\hat{k}%
_{i}$. To view this expression as diagonalization in $ii^{\prime }$ indexes,
we treat $\Delta _{S}(\hat{k}_{i})$ as eigenvectors and multiply the right
hand part by a set of eigenvalues $\varepsilon ^{(\kappa )}$ bearing in mind
that the physical solution for $\Delta _{S}^{(\kappa )}(\hat{k}_{i})$ is
given when the highest eigenvalue $\varepsilon ^{(\kappa )}$ becomes unity$.$
We thus obtain the Hermitian eigenvalue problem 
\begin{eqnarray}
&&\sum_{i^{\prime }}\left[ \sqrt{\frac{\delta A_{i}}{|v_{i}|}}M^{(S)}(\hat{k}%
_{i},\hat{k}_{i^{\prime }})\sqrt{\frac{\delta A_{i^{\prime }}}{|v_{i^{\prime
}}|}}-\delta _{ii^{\prime }}\frac{\varepsilon ^{(\kappa )}}{\ln \left( \frac{%
1.134\omega _{sf}}{T_{c}}\right) }\right] \times  \notag \\
&&\sqrt{\frac{\delta A_{i^{\prime }}}{|v_{i^{\prime }}|}}\Delta
_{S}^{(\kappa )}(\hat{k}_{i^{\prime }})=0,  \label{EQ-DIAG}
\end{eqnarray}%
where the renormalized eigenvalues $\lambda ^{(\kappa )}=$ $\varepsilon
^{(\kappa )}/\ln \left( \frac{1.134\omega _{sf}}{T_{c}}\right) $ come out as
a result of diagonalization. The condition $\varepsilon ^{(\kappa )}=1$ for
some $\kappa =m$ means that the highest renormalized eigenvalue $\lambda
^{(m)}=\max \{\lambda ^{(\kappa )}\}\equiv \lambda _{\max }$ is the physical
one which delivers $\Delta _{S}^{(m)}(\hat{k}_{i})$ and produces the famous
BCS equation for $T_{c}=1.134\omega _{sf}\exp (-1/\lambda _{\max }).$

We further modify the coupling constant that enters this equation to take
into account the discussed effects as in the electron--phonon theory. For
the mass enhancement, we introduce the Fermi surface (FS) average of the
electronic self--energy derivative taken at the Fermi level and define 
\begin{equation}
\lambda _{sf}=-\langle \frac{\partial \Sigma (\mathbf{k},\omega )}{\partial
\omega }|_{\omega =0}\rangle _{FS}  \label{Lsf}
\end{equation}%
For the Coulomb interaction operating at large energy scale, we introduce
the effective parameter $\mu _{m}^{\ast }$ which should now refer to the
same pairing symmetry $m$ as $\lambda _{\max }$. We therefore have the
effective coupling constant%
\begin{equation}
\lambda _{eff}=\frac{\lambda _{\max }-\mu _{m}^{\ast }}{1+\lambda _{sf}}
\label{Leff}
\end{equation}%
that should determine $T_{c}.$

\section{III. Results}

\subsection{a. Calculated Superconducting Properties in $HgBa_{2}CuO_{4}$}

Here we discuss the results of our calculated superconducting properties for
HgBa$_{2}$CuO$_{4}$ such as the energy gap function $\Delta _{S}(\mathbf{k}%
j) $ and the behavior of the maximum eigenvalue $\lambda _{\max }$
describing the strength of the spin fluctuational pairing. We use the full
potential linear muffin--tin orbital method \cite{FPLMTO} to calculate its
LDA energy bands and wave functions. The results show a rather simple band
structure near the Fermi surface composed primarily of the $d_{x^{2}-y^{2}}$
states of Cu hybridized with $O_{p_{x},p_{y}}$ orbitals on the square
lattice as is well known from the pioneering work of Emery\cite{Emery}. We
then utilize the LDA+FLEX(RPA) evaluation of the pairing interaction $%
K_{a_{1}a_{2}a_{3}a_{4}}^{(S)}(\mathbf{q})$ on the 20x20x4 grid of the $%
\mathbf{q}$ points in the Brillouin Zone (198 irreducible points). We use
Hubbard interaction parameter $U$ for the d--electrons of Cu as the input to
this simulation, which we vary between 2.5 and 4.5 eV. We also introduce the
doping by holes using the virtual crystal approximation.

The Fermi surface is triangularized onto small areas $\delta A_{i}$
described by about 1,600 Fermi surface momenta $k_{i}$ for which the matrix
elements of scattering between the Cooper pairs, $M^{(S)}(\hat{k}_{i},\hat{k}%
_{i^{\prime }}),$ are evaluated. The\ linearized BCS\ gap equation is then
exactly diagonalized and the set of eigenstates $\lambda ^{(\kappa )},\Delta
_{S}^{(\kappa )}(\hat{k}_{i})$ is obtained for both $S=0$ and $S=1$
pairings. The highest eigenvalue $\lambda ^{(m)}=\lambda _{\max }$
represents the physical solution and the eigenvector corresponds to
superconducting energy gap function $\Delta _{S}^{(m)}(\mathbf{k}j)$.

\begin{figure}[tbp]
\includegraphics[height=0.8152\textwidth,width=0.40%
\textwidth]{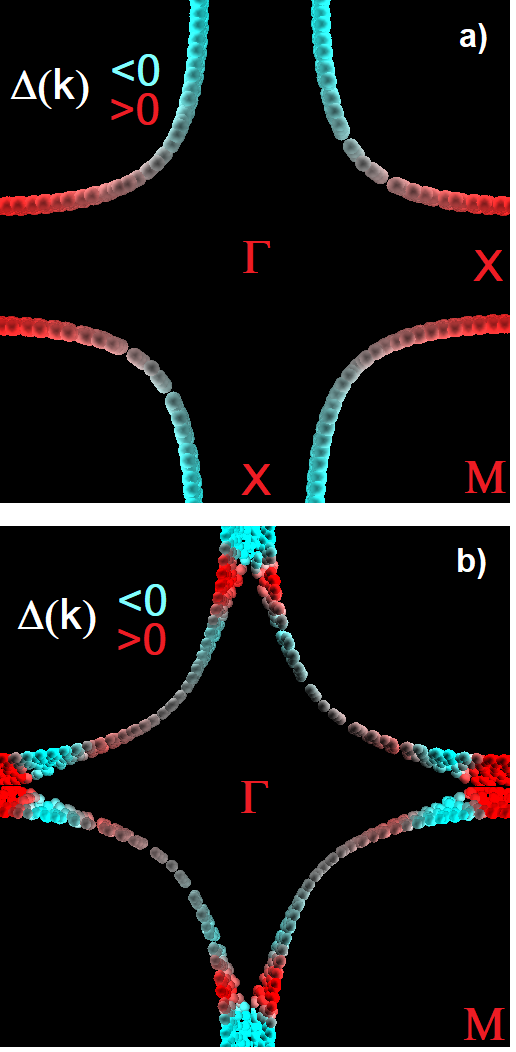}
\caption{{}Calculated superconducting energy gap $\Delta (\mathbf{k})$ for
singlet pairing in HgBa$_{2}$CuO$_{4}$ using numerical solution of the
linearized BCS gap equation with the pairing interaction evaluated using the
LDA+FLEX(RPA)\ approach described in text. Blue/red color corresponds to the
negative/positive values of $\Delta (\mathbf{k}).$ Plot a) is obtained for
doping $\protect\delta =0$ and shows typical d$_{x^{2}-y^{2}\text{ }}$%
behavior that is also seen for dopings $\protect\delta \leq 0.3$. Plot b)
corresponds to $\protect\delta =0.4$ and shows more oscillating behavior
highlighting the presence of higher--order harmonics.}
\label{FigDwaveGap}
\end{figure}

The result of this simulation is that $\Delta _{S=0}^{(m)}(\mathbf{k}j)$
shows a much celebrated d--wave behavior of $x^{2}-y^{2}$ symmetry (the
lobes pointing along $k_{x}$ and $k_{y}$ directions) This happens for
dopings $\delta \leq 0.3$ that we used in the simulation. A typical behavior
of this function is shown on Fig.\ref{FigDwaveGap}(a) for $U=4$ eV and $%
\delta =0$, where the blue/red color corresponds to negative/positive values
of $\Delta .$ The zeroes of the gap function are along (11) direction which
are colored in grey. This result is not surprising given the strong nesting
property of the Fermi surface around $\left( \pi ,\pi ,0\right) 2\pi /a$
point of the Brillouin Zone as was emphasized many times in the past.

We also studied the effect of higher dopings $\delta =0.4-0.5$. At those
values, the gap function retains the nodal lines along (11) but develops a
rather complex sign--changing behavior along the lobes by acquiring higher
order harmonics. We illustrate the solution in Fig.\ref{FigDwaveGap}(b) for $%
\delta =0.4$ and $U=3.3$ $eV.$ Such oscillatory behavior would carry an
additional kinetic energy and should be less favorable energetically.

\begin{figure}[tbp]
\includegraphics[height=0.39\textwidth,width=0.40\textwidth]{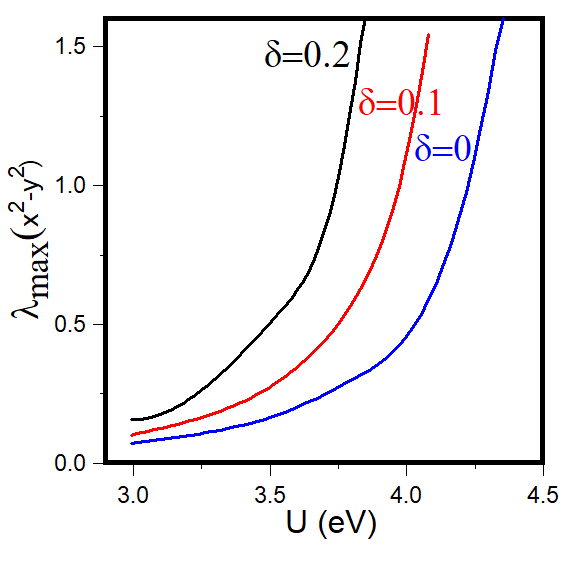}
\caption{{}Calculated dependence of maximum eigenvalue $\protect\lambda %
_{\max }$ of the linearized BCS equation as a function of the on--site
Hubbard interaction U for \textit{d}-electrons of Cu and for several hole
dopings $\protect\delta =0,0.1,0.2$ in HgBa$_{2}$CuO$_{4}$. Large values of $%
\protect\lambda _{\max }$ are seen for the values of $U$ close to the
antiferromagnetic instability.}
\label{FigLambdas}
\end{figure}

We further analyze the behavior of the highest eigenvalue $\lambda _{\max }$
as a function of $U$ and doping. The plot of $\lambda _{\max }$ vs. $U$ is
shown in Fig.\ref{FigLambdas} for hole dopings $\delta =0.0,0.1,0.2.$ In
particular, one can see pretty big $\lambda ^{\prime }s$ once we approach
the spin density wave instability for $U^{\prime }s$ close to 4 eV.
Unfortunately, this sensitivity imposes some challenges regarding the
predictions for $T_{c}$. It is however clear that if one adopts a constant $%
U $ value for all dopings, this plot will be incompatible with the
well--known dome--like behavior of the $T_{c}$ vs. doping. Rather, one need
to assume that for the undoped case $U$ is largest to trigger the
antiferromagnetic instability but then it gradually decreases with doping.

It is interesting to mention several works that do indeed see that $U$\
decreases with doping. A recent work\cite{cRPA} reported computation of
doping dependent $U$ using the constrained RPA (cRPA) procedure. Their
reported values of $U\approx $4 eV\ for HgBa$_{2}$CuO$_{4}$ are very close
to the ones needed to produce large $\lambda _{\max .},$ as seen in Fig. \ref%
{FigLambdas}, together with the trend that $U$ decreases with doping a
little bit. Another recent cRPA\ study reported this value to be 3eV for the
same compound\cite{cRPA2} which is again within the range of what we use in
our simulation. Unfortunately, the spread in these values also indicates
that we cannot rely on the present state--of--the art calculations of $U$.

In a different work employing DCA\cite{DCA-U}, an effective temperature
dependent coupling $\bar{U}(T)$ was introduced to parametrize the DCA\
pairing interaction in terms of the spin susceptibility. It was extracted
between 4 and 8 in the units of the nearest neighbor hopping $t.$ The latter
is known to be around 0.5 eV in the cuprates thus placing $\bar{U}(T)$
between 2 and 4 eV. $\bar{U}(T)$ has shown a significant reduction upon
doping.

At the lack of the accurate determination of $U$, we turn to more empirical
findings whether some other well--known properties of the cuprates can be
reproduced using our implemented LDA+FLEX(RPA)\ method. Those are related to
the normal state electronic structure and the extracted quasiparticle mass
enhancement, $m^{\ast }/m=1+\lambda _{sf}$. These data are needed to
determine the effective coupling constant, Eq. (\ref{Leff}) and give
estimates for the $T_{c}.$ They will be discussed below.

\subsection{b. Calculated Correlation Effects in Electronic Structure of $%
HgBa_{2}CuO_{4}$}

Here we discuss our calculated properties of HgBa$_{2}$CuO$_{4}$ in the
normal state. The self--energy is computed by utilizing procedure described
in Ref. \cite{LDA+FLEX} with full frequency resolved dynamical interaction
matrix $\hat{K},$ Eq. (\ref{EQ-K}). This is done as a "one--shot"
calculation using the Green functions obtained from the LDA\ band structure,
without feedback of the self--energy that will lead to "dressed" Green
functions. Due to the existence of a generating functional for FLEX
approximation\cite{FLEX}, the self--consistency with respect to the Green
functions can in principle be considered. There is one complication which
makes such implementation not straightforward and time consuming that once
complex self--energy is introduced, single--particle excitations are dumped
and no longer represented by the real energy bands $\epsilon _{\mathbf{k}j}.$
A general formulation via, for example, imaginary Matsubara frequencies is
needed. The effect of such self--consistency was studied earlier using the
GW method\cite{GW} with applications to some real materials\cite{SCF-GW}.
The outcome is that self--consistency worsens the agreement of the
one--electron spectra with experiment. Therefore the advantage of the
self--consistent implementation within perturbation theory is not obvious in
general.

Nevertheless, this issue may deserve a further investigation since cuprates
are doped Mott insulators in close proximity to the Mott transition and it
is known that in this regime, the self--consistency is an important step
when using, for example, dynamical mean field theory. Although not currently
implemented by us, one can adopt a simplified version of the
self--consistency with respect to quasiparticles, \textit{i.e.}, when not
the full self--energy but its value at $\omega =0$ and its frequency
derivative around $\omega =0$ describing the quasiparticle mass enhancement
are used to reconstruct new densities and resulting quasiparticle Green's
functions. It was developed in connection with the GW\ approach, and was
shown to reproduce the band gaps of semiconductors quite well \cite{QSGW}. A
combination of the LDA and Gutzwiller's method (called LDA+G) explores a
similar idea \cite{LDA+G} where the variational Gutzwiller method is used to
find those self--energy parameters. It was also implemented in a most recent
combination of the GW and DMFT called QSGW+DMFT\cite{QSGW+DMFT}.

\begin{figure}[tbp]
\includegraphics[height=0.783\textwidth,width=0.40%
\textwidth]{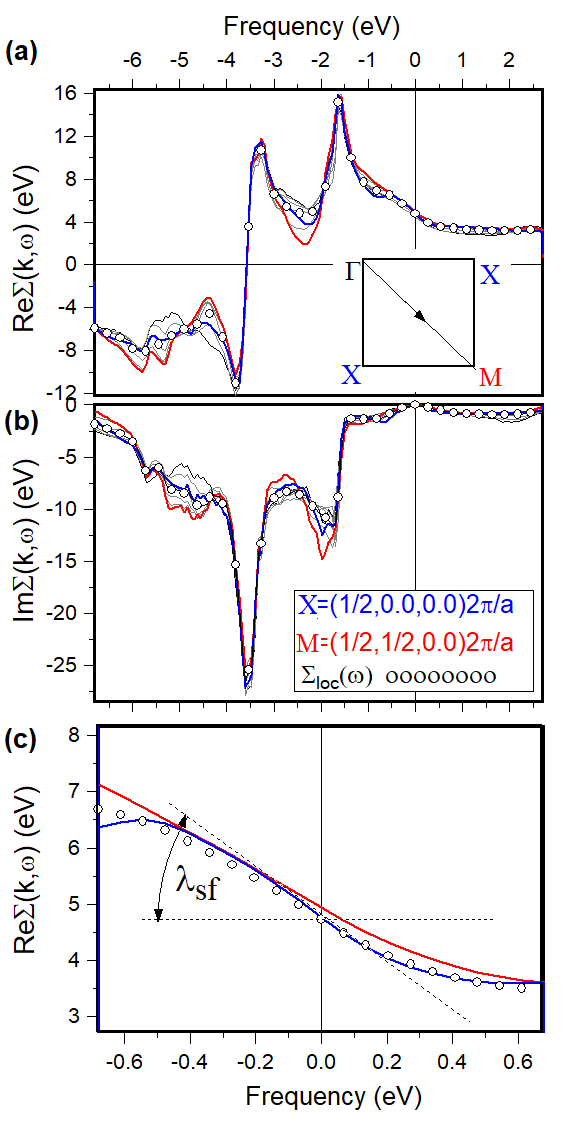}
\caption{{}Calculated d$_{x^{2}-y^{2}}$ diagonal matrix element of the
self--energy $\Sigma (\mathbf{k},\protect\omega )$ (a is the real part, b is
imaginary part, c is the real part at small energy scale) using FLEX--RPA
approximation for d electrons of Cu in HgBa$_{2}$CuO$_{4}$. Various black
curves show the self--energy for the wavevector k traversing along ($\protect%
\xi \protect\xi 0$) direction of the Brillouin Zone. Red/blue curves give
the result for the M/X points of the BZ. The circles show the result of the
local self--energy approximation taken as the average over all k--points. A
representative value of Hubbard $U$=4 eV is used and the doping $\protect%
\delta $ is set to zero in this plot, but similar trends are seen for a
whole range of U's and dopings studied in this work.}
\label{FigSelfEnergy}
\end{figure}

For the calculated Cu d--electron self--energy matrix $\Sigma _{a_{1}a_{2}}(%
\mathbf{k},\omega )$ , we found the only significant matrix elements of this
matrix exist for d$_{x^{2}-y^{2}}$ orbitals. This result is shown in Fig. %
\ref{FigSelfEnergy} where the diagonal matrix elements, \textrm{Re}$\Sigma (%
\mathbf{k},\omega ),$Fig. \ref{FigSelfEnergy}(a) and \textrm{Im}$\Sigma (%
\mathbf{k},\omega ),$\ Fig. \ref{FigSelfEnergy}(b), of $\Sigma _{a_{1}a_{2}}(%
\mathbf{k},\omega )$ with $a_{1}=a_{2}=x^{2}-y^{2}$ are plotted as a
function of frequency for several k points of the Brillouin Zone. A
representative value of $U$=4 eV and $\delta =0$ are used but general trends
of this function are similar for the range of $U^{\prime }$s and dopings
that we study here. The Hartree Fock value for $\text{Re}\Sigma $ has been
subtracted.

\begin{figure}[tbp]
\includegraphics[height=0.378\textwidth,width=0.40\textwidth]{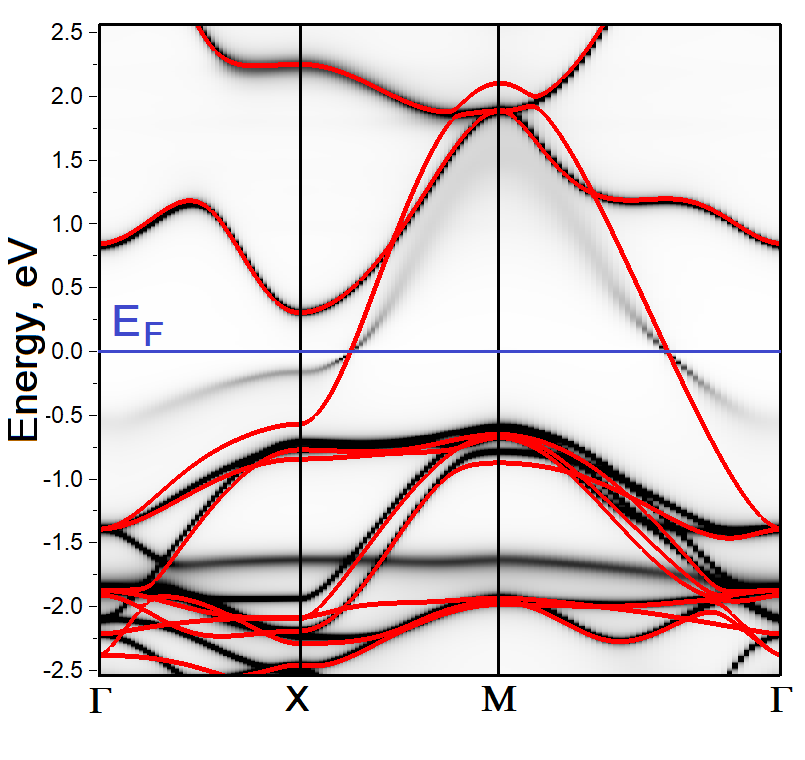}
\caption{{}Effect of the FLEX(RPA) self-energy on the calculated poles of
single particle Green's functions (shown in black) for undoped HgBa$_{2}$CuO$%
_{4}$ as compared with its nonmagnetic LDA\ band structure (red lines). The
local value at $\protect\omega =0$ is subtracted from $\Sigma (\mathbf{k},%
\protect\omega )$ during the calculation of $\text{Im}G(\mathbf{k},\protect%
\omega )\ $and the Hubbard $U$=4 eV is used.}
\label{FigBands}
\end{figure}

To illustrate the k--dependence, the self--energy is plotted in Fig.\ref%
{FigSelfEnergy} along $\Gamma M$ line of the Brillouin Zone (BZ) and also
for the $X$ point. At the energy scale -6eV<$\omega $%
<+2eV that we use in Fig.\ref{FigSelfEnergy}(a) and (b), we
find the k--dependence to be quite small prompting that the local
self--energy approximation may be adequate. This is not surprising since
within RPA, $\hat{\Sigma}=\hat{G}\hat{K}$ and the range of the self--energy
in real space is set by the interaction $\hat{K}$ which describes the
screening of the manifestly local $U$. In the k space, all features in $\hat{%
K}$ due to nesting come under the integral over the Brillouin Zone (BZ) and
averaged out.

We subsequently evaluate numerically the local self--energy $\Sigma
_{loc}(\omega )$ \ as an integral over all k--points. Its frequency
dependence is also shown in Fig. \ref{FigSelfEnergy} by small circles. We
see a close agreement between $\Sigma _{loc}(\omega )$ and $\Sigma (\mathbf{k%
},\omega )$.

\begin{figure}[tbp]
\includegraphics[height=0.4\textwidth,width=0.40\textwidth]{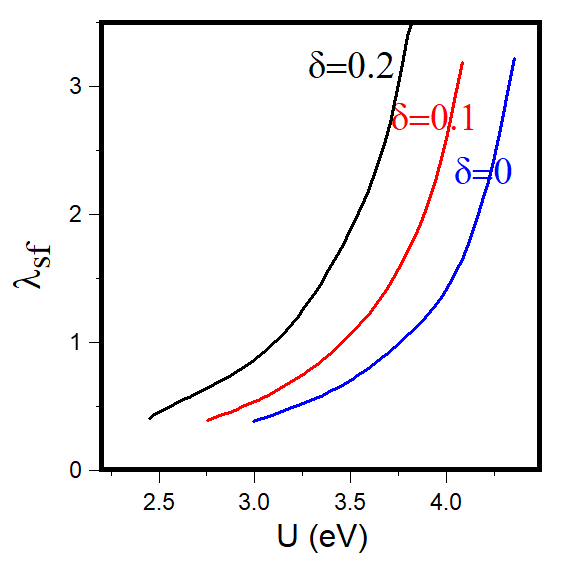}
\caption{{}Calculated dependence of the mass enhancement parameter $\protect%
\lambda _{sf}=m^{\ast }/_{m_{LDA}}-1$ as a function of the on--site Hubbard
interaction U for d-electrons of Cu and for several hole dopings $\protect%
\delta =0,0.1,0.2$ in HgBa$_{2}$CuO$_{4}$. Large values of $\protect\lambda %
_{sf}$ are seen for the values of $U$ close to the antiferromagnetic
instability.}
\label{FigLambdasSF}
\end{figure}

Another feature seen in this calculation is the development of pole like
behavior for the self--energy at frequencies around 2 and 4 eV. Those
resonances are frequently led to additional poles in the one--electron Green
functions that cannot be obtained using single--particle picture. The
imaginary part of the self--energy is quite large which indicates the
existence of strongly damped excitations. Those are usually hard to
associate with actual energy bands and detect by such experimental
techniques as ARPES which works best for the quasiparticles just below the
Fermi energy.

Fig.\ref{FigSelfEnergy}(c) shows the behavior of \textrm{Re}$\Sigma (\mathbf{%
k},\omega )$ on the scale $\pm 0.6$ eV with respect to the Fermi level for
the two representative points $M$ (red line)\ and $X$ (blue line) of the BZ
together with the momentum integrated self--energy (circles). A slight
variation in the slope of the self--energy at $\omega =0$ can be noticed as
well as some differences are seen in the frequency behavior. These data are
important for further analysis since the slope at $\omega =0$ sets the mass
enhancement parameter $\lambda _{sf}$ for the quasiparticles as illustrated
in Fig.\ref{FigSelfEnergy}(c)

Based on our calculated d--electron self--energy $\Sigma (\mathbf{k},\omega
) $, we evaluate the poles of the single particle Green function. The
obtained $\text{Im}G(\mathbf{k},\omega )$ for HgBa$_{2}$CuO$_{4}$ is plotted
in Fig. \ref{FigBands}. Most of the poles are seen as sharp resonances
(plotted in black) in the function $\text{Im}G(\mathbf{k},\omega )$ that
closely follows the energy band structure obtained by LDA plotted in red.
The notable difference is seen in the behavior of the hybridized Cu$%
_{d_{x^{2}-y^{2}}}-$O$_{p_{x},p_{y}}$ band in the vicinity of the Fermi
surface that acquires a strong damping at energies away from the Fermi
level. This is because our projectors allow the self--energy corrections for
the Cu d--electrons only. In order to generate the $\text{Im}G(\mathbf{k}%
,\omega )$ we have subtracted from $\Sigma (\mathbf{k},\omega )$ its local
value $\Sigma _{loc}(\omega )$ taken at $\omega =0$ which preserves the
shape of the Fermi surface as obtained by LDA. As one sees, the primary
effect of the self--energy is the renormalization of the electronic
bandwidth. The mass enhancement $m^{\ast }/m_{LDA}=1+\lambda _{sf}$ for the
Fermi electrons was found to be fairly k--independent. The value of $\lambda
_{sf}$ was calculated numerically as the average derivative of the
self--energy, Eq. (\ref{Lsf}), and estimated to be around 2.7 for $\delta =0$
and $U=$4 eV that we use in Fig. \ref{FigBands}.

We further analyze the dependence of $\lambda _{sf}$ on $U$ and doping$.$ It
was found to exhibit the behavior similar to the maximum eigenvalue $\lambda
_{\max }$ shown in Fig.\ref{FigLambdas}. To generate such functional
dependence we implement analytical differentiation of the self--energy at
zero frequency using spectral representation for the dynamically screened
interaction $K(\mathbf{q},\omega )$ proposed many years ago\cite{Winter}.
Fig \ref{FigLambdasSF} shows the calculated behavior of $\lambda _{sf}$ $\ $%
for dopings $\delta =0,0.1,0.2$ and 2.5 eV$<U<$4.5 eV. Although RPA\ does
not reproduce the metal--insulator transition, it signals its proximity via
the divergence of the quasiparticle mass as the system approaches the
instability.

We can compare the values of $\lambda _{sf}$ with the experimentally deduced
quasiparticle masses that were measured by ARPES experiments. There is some
spread in this data as reported in the past literature. Doping and
temperature dependence of the mass enhancement has been carefully studied
for Bi$_{2}$Sr$_{2}$CaCu$_{2}$O$_{8+\delta }$\cite{ARPES-MASS}, which
produced 0.5$\lesssim \lambda _{sf}\lesssim $1.7$.$ A later work\cite%
{ARPES-MASS2} for Bi$_{2}$Sr$_{2}$CaCu$_{2}$O$_{8}$ and also for La$_{2-x}$Ba%
$_{x}$CuO$_{4}$ reported the estimate $1\lesssim \lambda _{sf}\lesssim 2.$
Somewhat larger values of the self--energy slope, $4\div 8,$ taken for
several Fermi momenta have been seen in ARPES analysis of Bi$_{1.74}$Pb$%
_{0.38}$Sr$_{1.88}$CuO$_{6+\delta }$\cite{ARPES-MASS3}. The value of 2.7
along the nodal line was quoted for YBa$_{2}$Cu$_{3}$O$_{6.6}$\cite%
{ARPES-MASS4}. 
\begin{figure}[tbp]
\includegraphics[height=0.4\textwidth,width=0.40%
\textwidth]{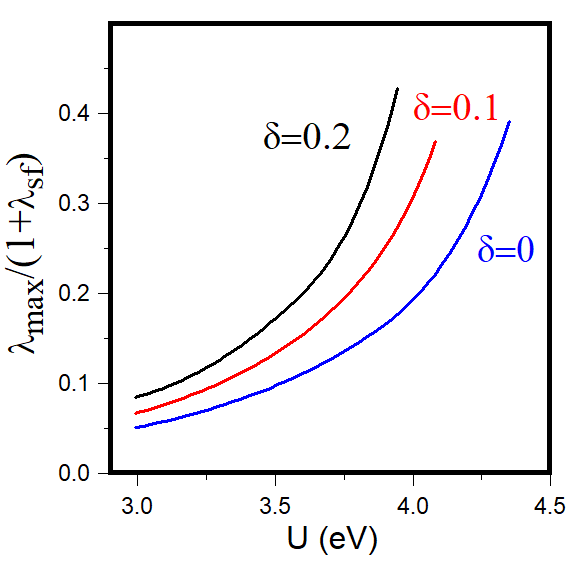}
\caption{{}Dependence of the effective spin fluctuational coupling constant $%
\protect\lambda _{eff}$ as a function of the on--site Hubbard interaction U
for d-electrons of Cu and for several hole dopings $\protect\delta %
=0,0.1,0.2 $ in HgBa$_{2}$CuO$_{4}$, calculated as the ratio between the
maximum eigenvalue $\protect\lambda _{\max }$ of the gap equation and the
quasiparticle mass enhancement $1+\protect\lambda _{sf}.$ }
\label{FigLambdasEFF}
\end{figure}

Quantum oscillations is another technique that gives the direct measure of
the effective masses. The reported $m^{\ast }$ range from 1.9 to 5 (in units
of the free electron mass) for various cuprates including the value of $%
2.45\pm 0.15$ for HgBa$_{2}$CuO$_{4+\delta }$\cite{MASS5}. As the LDA band
masses are not very different from the free electron masses, this indicates
that $1\lesssim \lambda _{sf}\lesssim 4$. Given the spread in these numbers,
it is clear that our calculations for $\lambda _{sf}$ shown in Fig \ref%
{FigLambdasSF} cover the range of the experimental data quite well.

\subsection{c. Effective Coupling Constant and Estimate for $T_{c}$.}

To give estimates for the effective coupling constant, $\lambda _{eff},$ Eq.(%
\ref{Leff}), we first notice that for the case of angular momentum $l=2$
relevant here, the Coulomb pseudopotential $\mu _{m}^{\ast }$ that projects
the screened Hubbard interaction on $d_{x^{2}-y^{2}}$ cubic harmonic is
expected to be very small \cite{Alexandrov}. We therefore set this parameter
to zero. The plot of $\lambda _{eff}=\lambda _{\max }/(1+\lambda _{sf})$ vs. 
$U$\ is shown in Fig \ref{FigLambdasEFF} for three dopings $\delta
=0.0,0.1,0.2.$ One can see that the range of these values is quite modest as
compared to both $\lambda _{\max }$ and $\lambda _{sf},$ primarily due to
the fact that the rise in the eigenvalue of the gap equation, Fig. \ref%
{FigLambdas}, is completely compensated by the renormalization effect of the
electronic self--energy$,$ Fig. \ref{FigLambdasSF}.

We can judge about the relevant range of $\lambda _{eff}$ and deduce
corresponding values of $T_{c}$ using the experimentally measured mass
enhancement data. Let, for example, take the middle value $\lambda _{sf}=2$.
From Fig. \ref{FigLambdasSF}, using the values of $U$ that produce $\lambda
_{sf}=2,$ we find the corresponding values of $\lambda _{eff}$ in the range $%
0.17\div 0.25$ in Fig \ref{FigLambdasEFF}, depending on doping. The BCS$\
T_{c}\approx \omega _{sf}\exp (-1/\lambda _{eff})=1\div 8K$ if one takes $%
\omega _{sf}=$40 meV. Once we get closer to the SDW instability, the
effective coupling increases to the values 0.4 as seen from Fig \ref%
{FigLambdasEFF}. It can go up even further by tuning $U$. If we consider $%
\lambda _{sf}=3$, corresponding to the higher values of the mass enhancement
seen experimentally, we find $0.26\lesssim \lambda _{eff}\lesssim 0.36$ from
Fig \ref{FigLambdasEFF} and the BCS$\ T_{c}\approx 10\div 30K.$ Given the
exponential sensitivity of the $T_{c},$ these estimates are certainly not
far away from 100K range for which $\lambda _{eff}\approx 0.6\div 0.7$ would
be desired.

We can comment on the numerous past publications devoted to the
self--consistent solution of the Eliashberg equation on imaginary Matsubara
axis using RPA--FLEX and single--orbital tight--binding band structures on
the square lattice. Unfortunately, due to the use of the imaginary
frequencies, in most cases the theory goes straight to $T_{c},$ and it is
hard to make direct comparisons to elucidate sources of possible
discrepancies. In the very earlier work\cite{Shimahara}, the authors found
trends very similar to ours regarding the $T_{c}$ using a general $%
t-t^{\prime }$ tight--binding model: The $T_{c}$ is very small and the spin
fluctuational superconductivity is strongly suppressed in the vicinity of
SDW due to the renormalization effect of the electronic self--energy, while
the attractive pairing was found to have a divergent character near the
instability. Later solution of the same model\cite{Arita2000} obtained $%
T_{c}=0.02t\approx 100K,$ for $U=4t=2eV$ and $\delta =0.15.$%
\begin{figure}[tbp]
\includegraphics[height=0.4\textwidth,width=0.40%
\textwidth]{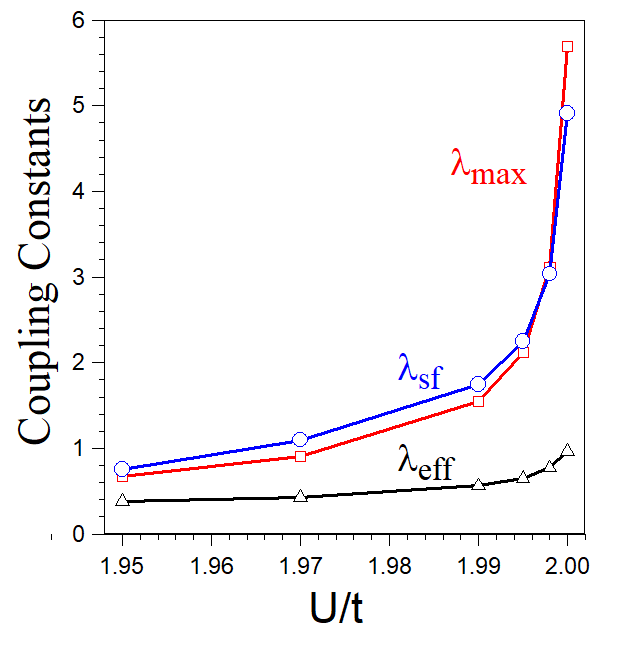}
\caption{{}Dependence of the maximum eigenvalue $\protect\lambda _{\max }$
of the linearized BCS equation (red line, suqares)$,$ the mass enhancement
parameter $\protect\lambda _{sf}$ $\ $(blue line, circiles) and the
effective spin fluctuational coupling constant $\protect\lambda _{eff}$ $=%
\protect\lambda _{\max }/(1+\protect\lambda _{sf})$ (black line, triangles)
as a function of the on--site Hubbard interaction U for the single-band
Hubbard model with $\protect\epsilon _{k}=-2t(\cos k_{x}a+\cos k_{y}a)$ and
doping $\protect\delta =0.1$ solved using the RPA(FLEX) method.}
\label{FigLambdas2DHUB}
\end{figure}

Given the last result, it is possible that our underestimation of $\lambda
_{eff}$ and $T_{c}$ is due to the fact that we utilize the full LDA energy
bands and the wave functions in the RPA--FLEX calculation. We have repeated
the procedure for the single--band tight--binding model $\epsilon
_{k}=-2t(\cos k_{x}a+\cos k_{y}a),$ and while seeing similar trends for both 
$\lambda _{\max }\ $and $\lambda _{sf}$ as a function of $U$, the evaluated $%
\lambda _{eff}$ as the ratio $\lambda _{\max }/(1+\lambda _{sf})$ is a
factor of two larger. We show the result of such caculation for $\delta =0.1$
in Fig \ref{FigLambdas2DHUB} close to the instability taking place right
above $U/t=2.$ Fixing $\lambda _{sf}=2(3),$ we extract $\lambda
_{eff}=0.62(0.78)$ and the BCS $T_{c}=90K(130K).$

A possible route for improving our approach would be to extend the BCS
approximation to include full frequency dependence of the pairing
interaction together with its implementation on the real frequency axis.
This should allow the comparison with the BCS limit in a more controllable
manner.

\section{IV. Conclusion.}

In conclusion, we have implemented the electronic structure calculation of
the superconducting pairing interaction using our recently developed
LDA+FLEX(RPA) method that accounts for the electronic self--energy of the
correlated electrons using a summation of the particle--hole bubble and
ladder diagrams. Based on this procedure, the superconducting scattering
matrix elements between the Cooper pairs have been evaluated numerically
which served as the input to numerical diagonalization of the linearized
BCS\ gap equation, whose maximum eigenvalue $\lambda _{\max \text{ }}$is
seen as the superconducting coupling constant$.$ The goal of this approach
was to establish the numerical procedure to evaluate material specific $%
\lambda $ without reliance on tight--binding approximations of the
electronic structure.

A case study of the prototype cuprate superconductor HgBa$_{2}$CuO$_{4}$ was
presented where we found a much celebrated d$-$wave ($x^{2}-y^{2}$ type)
symmetry of the superconducting energy gap as the favorable solution for the
whole range of dopings and on--site Hubbard interactions $U$ that were used
in our simulations. A strong dependence of $\lambda _{\max \text{ }}$ as a
function of $U$ was seen in the vicinity of antiferromagnetic instability.
The same was true for the calculated quasiparticle mass enhancement $m^{\ast
}/m_{LDA}=1+\lambda _{sf}$ in the normal state. The effective spin
fluctuational coupling constant $\lambda _{eff}=\lambda _{\max }/$($%
1+\lambda _{sf})$ was deduced, but found to be modest and incapable to
deliver high values of $T_{c}$ unless $U$ is tuned to be close to SDW.
Taking the experimental constraint for $\lambda _{sf}\lesssim 3$ we have
obtained the coupling constant $\lambda _{eff}\lesssim 0.4$ and the BCS\ $%
T_{c}\lesssim 30K.$ Application of the same procedure to the 2D Hubbard
model with nearest neighbor hoppings returns $\lambda _{eff}\approx 0.6\div
0.8$ and $T_{c}\approx 90\div 130K.$

At the end, we hope that with gaining further insights on other
unconventional superconductors using this approach and its further
improvements will ultimately allow us to reach a more quantitative
understanding of unconventional superconductivity in cuprates and other
systems.

\end{document}